\documentclass[apj]{emulateapj}
\input psfig.tex
\usepackage{amsmath}
\usepackage{subfigure}
\usepackage[]{graphicx}

\newcommand{\kms} {km/s}

\newcommand{\re} {R$_{\rm {e}}$}

\newcommand{\x} {$\chi^2$}
\newcommand{\mgb} {Mg\emph{b}}
\newcommand{\cote} {C{\^o}t{\'e} et al.}

\begin{document}

\shorttitle{The Rising Stellar Velocity Dispersion of M87 from Integrated Starlight}
\shortauthors{Murphy, Gebhardt \& Cradit}

\title{The Rising Stellar Velocity Dispersion of M87 from Integrated Starlight}

\author{Jeremy D. Murphy}
\affil{Department of Astrophysical Sciences, Princeton University, 4 Ivy Lane, Princeton, New Jersey 08544}

\author{Karl Gebhardt}
\affil{Department of Astronomy, University of Texas at Austin, 2515 Speedway C1400, Austin, Texas 78712}

\author{Mason Cradit}
\affil{Department of Physics, Southwestern University, P.O. Box 770, Georgetown, Texas, 78627} 

\email{jdm@astro.princeton.edu}

\begin{abstract}

We have measured the line-of-sight velocity distribution from integrated stellar light at two points in the outer halo of M87 (NGC~4486), the second-rank galaxy in the Virgo Cluster. The data were taken at R~=~480\arcsec\ ($\sim 41.5$~kpc) and R~=~526\arcsec\ ($\sim 45.5$~kpc) along the SE major axis. The second moment for a non-parametric estimate of the full velocity distribution is $420 \pm 23$~\kms\ and $577 \pm 35$~\kms\ respectively. There is intriguing evidence in the velocity profiles for two kinematically distinct stellar components at the position of our pointing. Under this assumption we employ a two-Gaussian decomposition and find the primary Gaussian having rest velocities equal to M87 (consistent with zero rotation) and second moments of $383 \pm 32$~\kms\ and $446 \pm 43$~\kms\ respectively. The asymmetry seen in the velocity profiles suggests that the stellar halo of M87 is not in a relaxed state and confuses a clean dynamical interpretation. That said, either measurement (full or two component model) shows a rising velocity dispersion at large radii, consistent with previous integrated light measurements, yet significantly higher than globular cluster measurements at comparable radial positions. These integrated light measurements at large radii, and the stark contrast they make to the measurements of other kinematic tracers, highlight the rich kinematic complexity of environments like the center of the Virgo Cluster and the need for caution when interpreting kinematic measurements from various dynamical tracers.

\end{abstract}


\section{Introduction}\label{sec:intro}

The assembly of stellar mass in elliptical galaxies has been the subject of significant investigation in recent years. With the discovery of a massive and old elliptical galaxy population at high redshift \citep[e.g.,][]{cim02a,cim04,mcc04a} we have been forced to revisit the mechanisms of massive galaxy growth. The puzzle is multifaceted. A central question is how such massive galaxies can exist in the early universe when they are generally considered the end-products of hierarchical assembly \citep[e.g.,][]{del07a}. Moreover, many of the high redshift quiescent galaxies appear to be particularly compact in comparison to local elliptical galaxies \citep[][]{dad05,tru06,tof07,van08b,cap09,cas10,wei11}, with half-light radii of $\sim 2$~kpc and no evidence for extended stellar halos \citep[e.g.,][]{szo12}. Further observations indicate that the very central regions (e.g., R $\le 1$~kpc) of z~$\approx 0$ galaxies are not physically denser than the high redshift population \citep[][although see Poggianti et al. 2013]{hop09,van10,tir11,van13,dul13} suggesting elliptical galaxy growth occurs predominantly at large radii \citep[e.g.,][]{van10}.

While there remains some debate about the degree of mass evolution in the cores of massive elliptical galaxies, the growth of mass at large radii (e.g., R $\ge 5$~kpc) is well established observationally \citep[][]{bui08,van08c,van10,new12,van13,pat13}. High spatial resolution simulations from cosmological initial conditions can recreate this mass growth at large radii, commonly through a growth history dominated by minor mergers \citep[][]{naa07,naa09,ose10,ose12}. However, both the degree of growth and how the mass is assembled over time remain poorly constrained observationally. In the case of very massive ellipticals and brightest cluster galaxies (BCG), the case is even more confounding as the BCGs at high redshift have been found to closely match the luminosity \citep[e.g.,][]{sto10}, mass \citep[][]{whi08,col09} and scale \citep[e.g.,][]{sto11} of BCGs in the local universe in certain work, but show significant growth in others \citep[][]{ber09,val10,asc11}. If we are to gain a complete understanding of how BCGs and other massive galaxies assemble their mass, we will need dynamical observations at large radii where the growth is expected to occur.

To this end we have made a measurement of the velocity dispersion of the integrated starlight in M87, the second-rank galaxy in the Virgo Cluster, using the Mitchell Spectrograph (formally VIRUS-P). This work is driven in part by the results of \citet{doh09} and \citet[][hereafter S11]{str11} who find a declining velocity dispersion profile from PNe and GC data respectively. Contrasting this are the results of \citet[][hereafter MGA11]{mur11} who found a rising velocity dispersion with radius for the stars. As certain systems have exhibited good agreement between different tracers of mass \citep[e.g.,][]{coc09} and others show disagreement \citep{rom01,chu10,she10,ric11}, we set out to make a measurement of the stellar velocity dispersion of M87 at a large radial distance in order to directly compare to the results of \citet{doh09} and S11. This approach was used by \citet{wei09} to good effect on the massive local ellipticals NGC~3379 and NGC~821.

The traditional estimate of M87's half-light radius (\re) from RC3 \citep{dev91} is $\sim 95$\arcsec. This estimate of the \re\ of M87 puts our measurements at $\sim 5.0$~\re\ and $\sim 5.5$~\re. However, \citet{kor09} measure an \re\ for M87 that is more than a factor of 7 times larger than the canonical value. A consideration of the deep photometry of \citet{mih05} and \citet{jan10} makes it clear that the definition of the half-light radius in the centers of clusters, particularly ones as unrelaxed as Virgo, is perhaps ill-defined. We therefore elect to not use \re\ in this work and will cite physical and/or on-sky scales where appropriate.

To remain consistent with our previous papers on M87 \citep[][]{geb09,geb11,mur11} we assume a distance to M87 of 17.9~Mpc. This distance corresponds to a scale of 86.5~pc~arcsec$^{-1}$. All references to other values in the literature are scaled accordingly. 

The paper outline is as follows. In \S~\ref{sec:data} we describe the observations and data collection. \S~\ref{sec:reductions} outlines the data reduction steps. Our results are found in \S~\ref{sec:results} where we also place our results in context with several velocity dispersion values from the literature. A discussion of the implications of these results in the context of both the assembly of mass in massive galaxies and comparisons of the different tracers of mass is found in \S~\ref{sec:discussion}. We summarize our results in \S~\ref{sec:summary}.

\vspace{1cm}
\section{Observations and Data Collection}\label{sec:data}

\begin{figure}
  \begin{center}
    \psfig{file=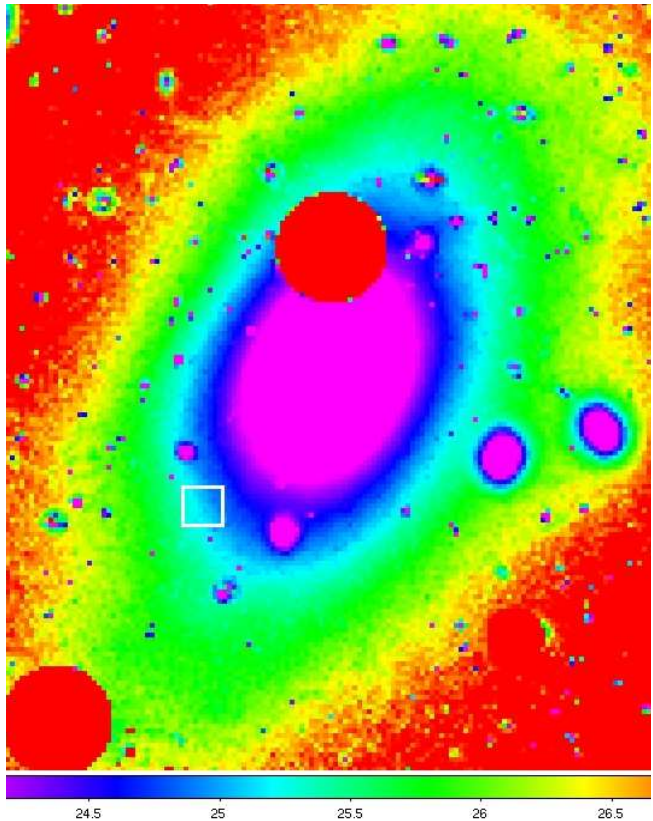,width=8.55cm,bbllx=54bp,bblly=10bp,bburx=588bp,bbury=720bp,clip=}
    \vskip10pt
    \figcaption{The deep photometry from \citet{mih05} showing the extended stellar halo of M87. The red circular regions have been masked in order to highlight the faint structure of the diffuse halo. North is up and West is to the right. The Mitchell Spectrograph field is shown as a white box to the SE. Our two fields are divided roughly along the diagonal of the field-of-view, from upper-left to lower-right. More details of the two fields can be found in Figure \ref{fig:fibflux}. The surface brightness (V) drops from $\sim 24.8$ to $\sim 25.4$ over our field \citep{kor09}. NGC~4486a is visible just to the lower-right of our field. \citet{pru11} report a recession velocity of 757~\kms\ which is significantly different than the previous value of 150~\kms\ \citep{weg03,rin08}. IC~3443 is seen above and slightly to the left of our field. IC~3443 exhibits significantly higher scatter in the literature values reported for its recession velocity, ranging from 1019~\kms\ \citep{evs07} to 2272~\kms\ \citep{dev91}, yet \citet{ade06} and \citet{rin08} have it at 1784 and 1785~\kms\ respectively.  Note the extended and distorted photometry of M87 along the major axis to the SE as compared to the NW. This distortion is more clearly seen in the photometry of \citet[][see Figure 4]{rud10}, and particularly in \citet[][see Figure 1]{wei97}.
      \label{fig:mihos}}
    \vskip-10pt
  \end{center}
\end{figure}

\subsection{The Mitchell Spectrograph}\label{sec:instrument}

The George and Cynthia Mitchell Spectrograph \citep[The Mitchell Spectrograph, formally VIRUS-P][]{hil08a} is currently deployed on the Harlan J. Smith 2.7~m telescope at McDonald Observatory. The Mitchell Spectrograph is a gimbal-mounted integral-field unit (IFU) spectrograph composed of 246 optical fibers each with a 4.2\arcsec\ on-sky diameter. The CCD is a $2048~\times~2048$ back-illuminated Fairchild 3041 detector. The wavelength range for these observations is 3550--5850~\AA\ at a resolution of R~$\approx 850$. This gives a spectral resolution of 4.7~\AA\ (FWHM, 146~\kms) at 4100~\AA\ and 4.95~\AA\ (117~\kms) at 5400~\AA. The fibers are laid out in an hexagonal array, similar to Densepak \citep{bar88}, with a one-third fill factor and a large ($107\arcsec \times 107\arcsec$) field-of-view. A detailed figure of this layout can be found in MGA11 (see Figure 2). See also the small insert in Figure \ref{fig:fibflux}. 

\subsection{Data Collection}\label{sec:datacol}

The Mitchell Spectrograph data presented here was collected over five partial nights in May 2010 and two complete nights in March 2011. From these observing runs 42 science frames were collected, each 30 minutes in duration and bracketed by 10 minute sky nods. The conditions for 2 of the 7 nights were photometric, with the conditions for the remaining 5 nights being good. At this low galaxy surface brightness ($\sim 24.8$ to $\sim 25.4$ in V across our field, Kormendy et al. 2009), any light cirrus strongly affects the quality of the frame. Therefore, during parts of the night with any noticeable cloud cover we observed other targets. In total, 38 of the 42 frames collected were clean enough to include in the final spectra presented here. The median seeing was 2.2\arcsec. This is well below the 4.2\arcsec\ on-sky fiber diameter and therefore does not influence our results.

The position of our science pointing is shown in Figure \ref{fig:mihos} overlaid on the deep photometry of \citet{mih05}. North is up and West is to the right. The Mitchell Spectrograph field is shown as the white box to the lower-left of the galaxy. Note the elongated shape and disturbed photometry to the SE along the major axis. This elongation is also seen clearly in \citet[][see Figure 1]{wei97} and \citet[][see Figure 4]{rud10}.

The Mitchell Spectrograph has no dedicated sky fibers, so nodding for sky frames is necessary and constitutes approximately one-third of our observing time. The sky nods were taken $\sim 1$ degree to the NNE of our science pointing, well away from the diffuse light in the center of the Virgo Cluster \citep{mih05,jan10}. As sky nods sample the sky at a different time from the science frames we must be careful to understand how much this influences our final kinematics. We return to an analysis of possible systematics due to our sky subtraction in \S~\ref{sec:skyuncertain}.

\section{Data Reduction}\label{sec:reductions}

The primary data reduction steps are completed with Vaccine, an in-house data reduction pipeline developed by Joshua J. Adams for the handling of Mitchell Spectrograph IFU data. We refer the interested reader to \citet{ada11a} where further details of Vaccine are given. In MGA11 we give the complete details of our data reduction methods, and provide a brief outline of the steps here.

\subsection{Vaccine}\label{sec:vaccine}

To begin, the bias frames for the entire observing run are combined, and all of the science, sky and calibration frames are overscan and bias subtracted. Neighboring sky nods for each science frame are appropriately scaled and combined (see \S~\ref{sec:sky} for further details on this step). These built-up sky nods are then taken through the rest of the reductions in the same manner as the science frames. The arcs and twilight flats for a single night are then combined. Curvature along both the spatial and spectral direction of the CCD is accounted for by considering each fiber individually during the reduction process. To account for curvature in the spatial direction, a $4^{\text{th}}$ order polynomial is fit to the peaks of each of the 246 fibers of the twilight flat frames taken each night. The 246 polynomial fits are then used on each science and sky frame to extract the spectra, fiber by fiber, within a 5 pixel-wide aperture. The pixel values are extracted directly, without interpolation of any kind. This allows us to avoid correlation in the noise and track the exact noise in each pixel throughout the Vaccine reductions.

To address curvature in the spectral direction, a wavelength solution is determined for each fiber and each night. To accomplish this, a $4^{\text{th}}$ order polynomial is fit to the centers of known mercury and cadmium arc lamp lines, fiber by fiber, from the calibration frames. This polynomial fit then becomes the unique wavelength solution for a given fiber on a given night.

Next, the twilight flats are ``normalized'' to remove the solar spectrum following a procedure similar to the sky modeling method in \citet{kel03}. Further details of this method are given in \citet{ada11a}. In brief, this process creates a model of the twilight sky free of flat-field effects. This sky model is then divided out of the original flat to remove the solar spectrum while preserving the flat-field effects we want to capture (namely, pixel-to-pixel variation, fiber-to-fiber throughput, and fiber cross-dispersion profile shape). These normalized flats are then used to flatten the science and sky data. Once the built-up sky frames are flattened, a sky model is made in the same manner as the flat-field. This model of the sky is then subtracted from the science frames. As accurate sky subtraction is critical to this work, we give more complete details of how the sky subtraction is handled in \S~\ref{sec:sky}. Finally, cosmic rays are located and masked from each 30 minute science frame.

The low galaxy surface brightness over our pointing \citep[$\sim 24.8$ to $\sim 25.4$~mag~arcsec$^{-1}$ in V;][]{kor09} requires us to co-add many fibers to reach the requisite signal-to-noise (S/N). To accomplish this, we identify all fibers that do not fall on foreground or background objects. After rejecting fibers that show evidence for extra light we are left with 223 of the original 246 fibers. These fibers are then divided into two spatial bins by their radial position from the center of M87. The first radial bin (R1) contains 99 fibers and has a corresponding S/N of 28.9. The second bin (R2) contains 114 and a S/N of 25.2. The radial range for the R1 pointing spans 437.2\arcsec\ to 506.4\arcsec\ with a luminosity-weighted center of $\sim 480$\arcsec. For the R2 pointing the radial range covers 506.8\arcsec\ to 562.2\arcsec\ with a luminosity-weighted center of $\sim 526$\arcsec. Further details of the fiber selection are given in \S~\ref{sec:split}. All the individual spectra are interpolated onto a common wavelength scale before being combined via the biweight \citep{bee90}.

\subsection{Sky Subtraction}\label{sec:sky}

Working $\sim 3$ magnitudes below the night sky involves a close accounting of how sky subtraction is handled. The sky nods necessarily sample the sky at a different point in time, and cost observing time to collect, yet are necessary at this faint surface brightness and with a galaxy this large on sky. A complete description of our method of sky subtraction is given in the appendix of MGA11. Here we provide an overview of our method of sky subtraction, and some specifics unique to these data.

To summarize our sky subtraction procedure, each 10 minute sky nod gets scaled by a factor of 1.5, then coadded to the scaled sky nod that brackets the science frame. This scaling and coadd of the sky nods leads to an equivalent exposure time to the science frames. A range of sky nod scaling, varying in $\pm 2\%$ increments, are explored in order to both quantify how slight over and under subtraction of the sky affects our final values, and to account for possible changes in transparency over the 30 minute science exposure. All of the scaled and coadded sky nods are then taken through the Vaccine reductions as described in \S~\ref{sec:vaccine}. A model of the night sky in each fiber is made via a modification of the sky-subtraction routine of \citet{kel03}. This model is subtracted directly from the science frame. The range of sky frame scalings leads to a range of sky-subtracted science frames. The determination of the best subtraction is made by a visual inspection of both night sky lines and absorption features in each science frame. As any slight deterioration in conditions led us to turn to brighter targets, in nearly all cases a scaling of 1.5 to each sky nod returned the best sky subtraction. The various scalings that are not considered optimal are still carried through the complete set of reductions and used to quantify uncertainties as discussed in \S~\ref{sec:skyuncertain}.

\subsection{Extracting the Stellar Kinematics}\label{sec:extract}

\begin{figure*}[ht]
  \begin{center}
    \centerline{
\psfig{file=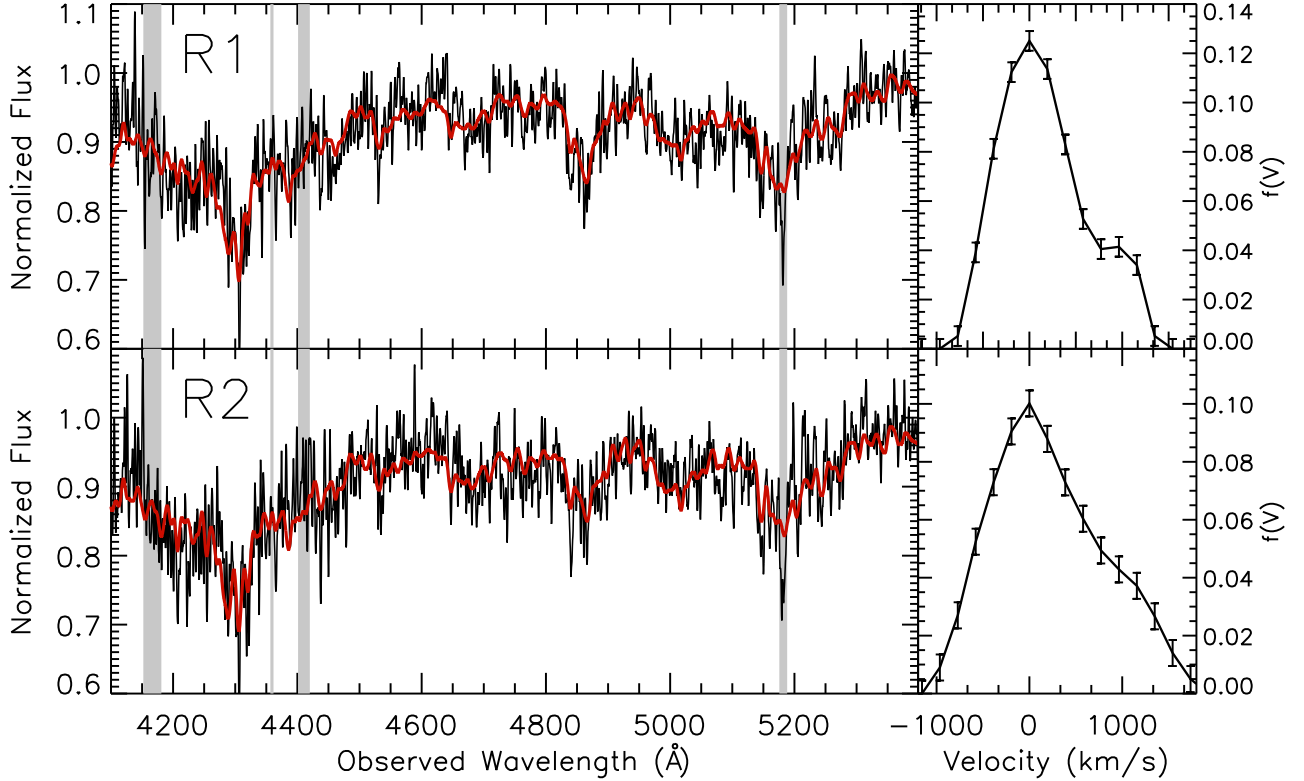,width=18.5cm,bbllx=54bp,bblly=410bp,bburx=558bp,bbury=720bp,clip=}}   
    \figcaption{The spectra (black line) for the R~$\sim 480$\arcsec\ (top: R1) and R~$\sim 526$\arcsec\ (bottom: R2) fields on M87. The red line plots the best-fit set of template stars convolved with a line-of-sight velocity distribution profile (LOSVD), shown to the right of each spectrum. The regions marked by vertical gray bands have been suppressed from the template fitting due to issues with the subtraction of night sky lines. The resulting LOSVDs are shown with uncertainties as determined by 100 Monte Carlo realizations of the kinematic extraction, described in \S~\ref{sec:mc}.
    \label{fig:spec}}
    \vskip-10pt
  \end{center}
\end{figure*}

The final line-of-sight velocity distribution (LOSVD) is effectively a histogram showing the distribution of stellar velocities along our line-of-sight. In order to determine the LOSVD profile from our spectra we apply the maximum penalized likelihood (MPL) method described in \citet{geb00b} and \citet{pin03}. The MPL technique is based upon the methods detailed in \citet{sah94} and \citet{mer97}, yet differs from their work in that we fit for the velocity bins and template weights simultaneously. This procedure was shown in \citet{pin03} to give similar results to the Fourier correlation quotient technique of \citet{ben90}. We will use the terms ``LOSVD'' and ``velocity profile'' interchangeably.

To determine the best LOSVD the following steps are taken: 1) An initial guess at a velocity distribution profile is made. The profile is divided into 29 velocity bins. The total area (e.g., height) given to each velocity bin is allowed to vary, yet the total of all bins must sum to unity. 2) A set of 16 stellar templates are used (see Table \ref{table:tlist}). The weight given to each template star is allowed to vary between 0.0 and 1.0. The total weight of all template stars must sum to unity. 3) The composite template star is convolved with the LOSVD and fit to the spectrum. 4) The heights of each velocity bin, and the weights given to each template star, are allowed to vary simultaneously until the residuals between the convolved, composite template spectrum and the observed spectrum are minimized. Smoothness is imposed on the LOSVD by the addition of a penalty function to the final \x\ of the fit \citep{mer97}.

In Figure \ref{fig:spec} we plot the results of this process. The two spectra shown (black lines) are for the R1 (top) and R2 (bottom) fields of our Mitchell Spectrograph pointing. The red line is the best-fit composite template star, convolved with the LOSVD. The vertical gray regions in the figure denote spectral regions withheld from the kinematic extraction due to issues with poor subtraction of night sky lines. The best-fit LOSVD, with uncertainties based on Monte Carlo simulations (see \S~\ref{sec:uncertainties}), is shown to the right of each spectrum.
The most notable feature in both of these velocity profiles is the strong asymmetry to positive velocities. The robustness of the asymmetry seen in both fields, and how we have elected to fit to the asymmetry, is explored in \S~\ref{sec:tests} and \S~\ref{sec:fitting}.

\vskip20pt

\subsubsection{Stellar Template Library}\label{sec:library}

The stars used in the kinematic extraction were taken from the Indo-US spectral library \citep{val04}. Initially a set of 40 template stars were explored and iteratively rejected down to the final list presented in Table \ref{table:tlist} based on whether the star received any weight in the extraction process. The final set of 16 template stars are not the same as those used in MGA11. This choice was made to cover a broader range in stellar type and metallicity. We have checked the extraction against the set of template stars used in MGA11 and see no significant differences in the resulting kinematics. In both MGA11 and Murphy et al. (in preparation) we explore systematics in the choice of the template library by comparing the first 4 moments of our extracted LOSVD when using the same set of template stars, yet taken from the MILES stellar library \citep{san06b}. A small ($\le 7$~\kms) difference in velocity is the only systematic offset seen between the two template libraries. We elect to use the Indo-US library for all the work shown here.
  
At our radial position on M87 the surface brightness ranges between  $\sim 24.8$ and $\sim 25.4$ in V \citep{kor09} which is a factor of $\ge 20$ below the night sky. Therefore, the S/N of the resulting spectra is not high enough to break the Mitchell Spectrograph's 3550 -- 5850~\AA\ wavelength range into multiple regions for kinematic extraction as done in MGA11. After much iteration, we have elected to fit the spectra between 4100--5400~\AA, rejecting the Ca H\&K region and redward of the \mgb\ spectral feature. 

\begin{deluxetable}{lllr}
  \tablewidth{235pt}
  \tablecaption{Indo-US and Miles Template Stars
    \label{table:tlist}}
  \tablehead{
    \colhead{ID} & \colhead{Type} & \colhead{V} & \colhead{[Fe/H]}}
  \startdata
            &         &        &        \\
  HD 27295  &  B9IV   &  5.49  &  -0.74 \\
  HD 198001 &  A1V    &  3.77  &   0.07 \\
  HD 2628   &  A7III  &  5.22  &   0.00 \\
  HD 89254  &  F2III  &  5.25  &   0.25 \\
  HD 9826   &  F8V    &  4.10  &   0.11 \\
  HD 136064 &  F9IV   &  5.10  &  -0.04 \\
  HD 33256  &  F2V    &  5.13  &  -0.33 \\
  HD 34411  &  G0V    &  4.70  &   0.06 \\
  HD 201889 &  G1V    &  8.04  &  -0.95 \\
  HD 3546   &  G5III  &  4.37  &  -0.66 \\
  HD 74462  &  G5IV   &  8.69  &  -1.41 \\
  HD 6203   &  K0III  &  5.41  &  -0.29 \\
  HD 9408   &  K0III  &  4.69  &  -0.30 \\
  HD 10780  &  K0V    &  5.63  &   0.10 \\
  HD 6497   &  K2III  &  6.42  &   0.01 \\
  HD 200527 &  M31bII &  6.27  &   0.70 \\
  \tablecomments{The template stars used in the determination of the best-fit LOSVD (Figure \ref{fig:spec}). These 16 stars were selected iteratively from an initial list of 40 Indo-US template stars based on whether the star received any weight in the fitting process. The same set of stars were taken from both the Indo-US and Miles stellar library and the final kinematics (i.e. the first 4 Gauss-Hermite moments) were compared and agreed within their uncertainties. A $\le 7$~\kms\ offset in velocity was seen between the two libraries, but no other significant systematics were seen.}
\end{deluxetable}

\subsubsection{Testing the Velocity Profile Asymmetry}\label{sec:tests}

The strong asymmetry in the extracted LOSVDs is a striking feature that was not seen in any of the velocity profiles at smaller radii in MGA11. We have tested the robustness of this asymmetry in a variety of ways which we discuss in turn here. 

The first natural test is to explore a range of the smoothing parameter discussed in \S~\ref{sec:extract} and detailed in \citet{mer97}. Generically, the smoothing parameter is a penalty placed on the distribution of velocities that deviates from the true distribution. Without prior knowledge of the shape of the true distribution oftentimes penalized methods return unphysical distributions. Specifically, we encorporate the penalized likelihood method of \citet{mer97} which assumes a Gaussian distribution as a prior. Therefore, as the penalty (i.e. smoothing) is turned up, the velocity distribution is driven to a Gaussian. This approach is preferred for noisy data as reduced S/N is commonly best handled by parameterizing the velocity distribution in some fashion.
  
Increasing the smoothing will naturally reduce the significance of the asymmetry as power in the individual velocity bins gets averaged over a wider range and driven towards a pure Gaussian. In MGA11 the smoothing was set by comparing the smoothing value with the second moment of the velocity profile. The goal is to apply the lowest level of smoothing the S/N of the spectra will allow. In high S/N spectra the smoothing parameter can take on a fairly wide range without altering the velocity profile significantly. As the S/N of the spectra decreases, the range of acceptable smoothing parameters typically narrows.

To explore how the smoothing influences the second moments in the R1 and R2 fields we have run a set of extractions with a range of smoothing values. At low levels of smoothing (0.001) the resulting velocity profile is noisier, yet quantitatively very similar to the profiles created with larger smoothing parameters. At smoothing values $\ge 1$, the asymmetry begins to blend with the rest of the velocity profile, driving the second moment steadily upward. Yet for smoothing values over the range of 0.001 to 0.5 the second moment deviates slightly ($\sigma_{\textrm{R1}} = 2.8$~\kms\ and $\sigma_{\textrm{R2}} = 27.4$~\kms) and non-systematically.

A second simple test of the LOSVD asymmetry is to combine the spectra from the R1 and R2 fields. Combining the data boosts the S/N and so should help to determine whether the asymmetry is a product of noisy spectra. While the S/N does not increase substantially when the R1 and R2 spectra are combined, the asymmetry clearly remains.

As the data were taken over two observing runs separated by nearly a year, we can test the asymmetry in a third way. We do this by splitting the data between the May 2010 and March 2011 observing runs. If the asymmetry stems from a systematic calibration error in one of the two observing runs (e.g., an error in the wavelength solution) it would likely show up in this test. The asymmetry remains clearly present in both the May 2010 and March 2011 velocity profiles. 

A fourth test of the LOSVD asymmetry comes from limiting the choice of template stars used in the kinematic extraction. For the fits shown in Figure \ref{fig:spec} the entire list of 16 template stars from Table \ref{table:tlist} were used. To explore whether the asymmetry is driven by the choice of template stars we ran the LOSVD extraction routine with a single template star (HD92588). Although template mismatch leads to a noisier LOSVD, the asymmetry remained clearly present.

A final test was conducted, with the assistance of Genevieve Graves, in order to rule out both the fitting procedure and template library as the source of the asymmetry. We ran the R1 and R2 spectra through pPXF \citep[][]{cap04} and used single burst stellar population models from \citet[][]{vaz10}, based on the empirical MILES stellar library \citep[][]{san06b}. The templates cover a range of ages (63 Myr to 17.8 Gyr) and metallicity ($-1.3 <$ [Z/H] $< 0.22$). The bias parameter in pPXF was left on, which penalizes fits with a non-zero $h_3$ and $h_4$, in an attempt to minimize the asymmetry. A comparison of the first four moments ($v$, $\sigma$, $h_3$ and $h_4$) of this kinematic extraction strongly support the asymmetry, with R1 field values of $v=-30.0$~\kms, $\sigma=593.7$~\kms, $h_3=0.235$ and $h_4=0.199$ and R2 values of $v=-53.4$~\kms, $\sigma=662.8$~\kms, $h_3=0.243$ and $h_4=0.293$.

Lastly, and perhaps most significantly, the velocity profiles of the ultra compact dwarfs (UCD), bright GCs (i$_0 < 20$) and blue GCs from \citet[][see Figure 23 in that work]{str11} at the same radial position show a very similar asymmetry to positive velocity. In \S~\ref{sec:discussion} we will argue, albeit tentatively, that both the evidence for the continued assembly of M87, and the velocity profile asymmetry seen in S11, supports the presence of a cooler and strongly offset stellar component. We will return to a comparison of the S11 results and this work in \S~\ref{sec:conflict}.

\subsubsection{Fitting the LOSVD}\label{sec:fitting}

\begin{figure*}[ht]
  \begin{center}
  \vspace{0.22cm}
    \centerline{
      \psfig{file=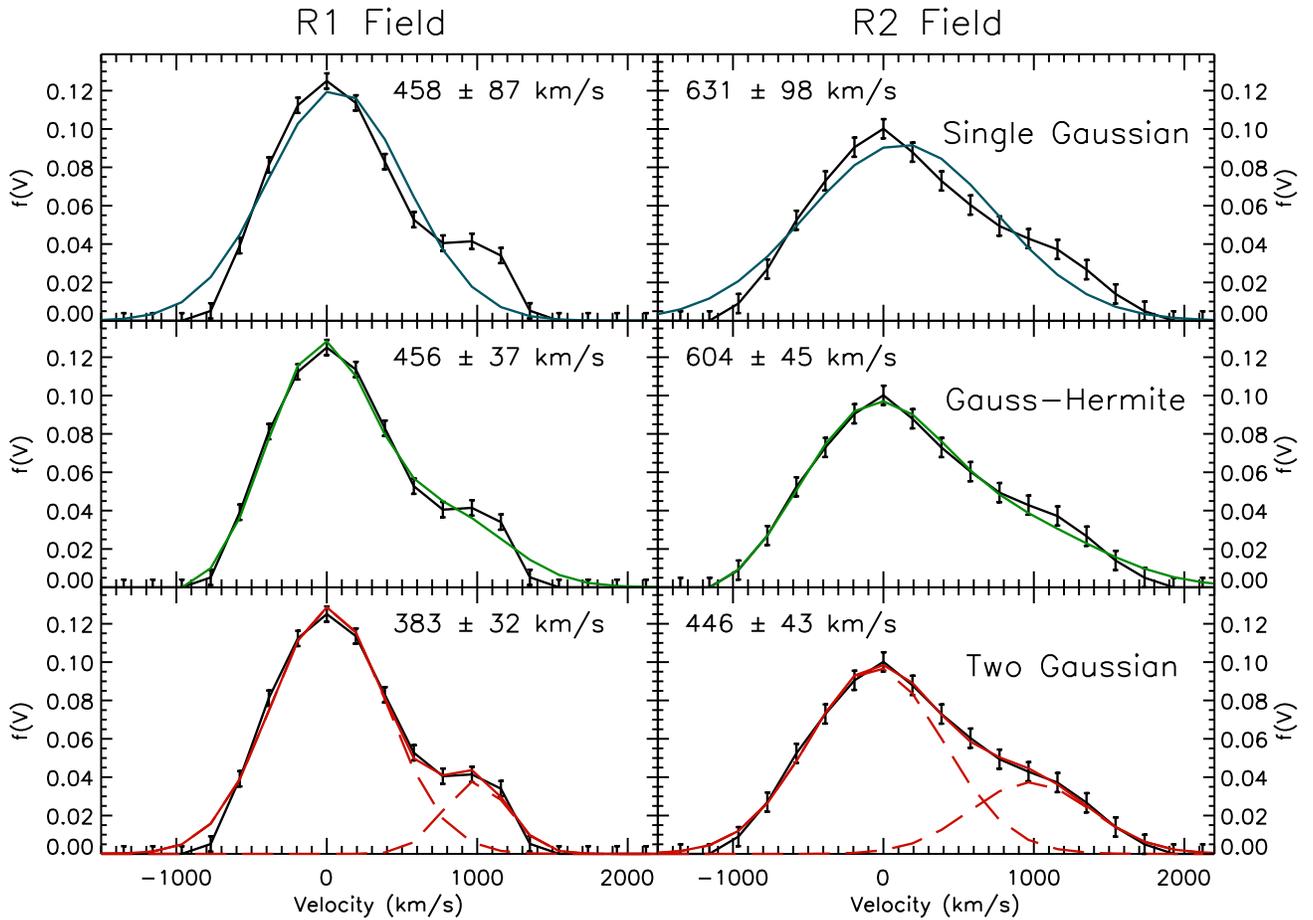,width=18.5cm,bbllx=54bp,bblly=360bp,bburx=558bp,bbury=720bp,clip=}}   
    \figcaption{The results for the single Gaussian (top, blue), Gauss-Hermite (middle, green) and two-Gaussian (bottom, red) parameterization of the R1 (left) and R2 (right) velocity profiles. The measured stellar velocity dispersion is written in each panel. The values for the Gauss-Hermite fits ($v$, $\sigma$, $h_3$ and $h_4$) can be found in the text. For the two-Gaussian fit, the dashed (red) lines plot the individual Gaussians, and the solid (red) line plots the sum of the two Gaussian components. Table \ref{table:2gauss} provides the percentage of the total area for each component, the component's center and second moment for the two-Gaussian fits.
    \label{fig:fits}}
    \vskip-10pt
  \end{center}
\end{figure*}

In \S \ref{sec:tests} we explored and ruled out a number of possible systematic effects that might give rise to the asymmetry to positive velocities seen in both the R1 and R2 LOSVDs. However, as the data are fairly noisy, we can not claim to have ruled out all possible systematics that could lead to the asymmetry. For this reason we explore various methods for fitting the velocity profiles and extracting the kinematics which explore the upper and lower limits of the stellar velocity dispersion.

The most direct way to measure the stellar velocity dispersion of M87 is to take the non-parametric second moment of the LOSVD. This measurement returns values of $420 \pm 23$~\kms\ and $577 \pm 35$~\kms\ for the R1 and R2 fields. This steep rise in the second moment occurs over $\sim 4.3$~kpc and, while quite striking, is not unprecedented as \citet[][]{ven10} see a similar rise in NGC~3311, the BCG in Abell~1060. As a non-parametric measure of the velocity profile is what matters for a dynamical analysis, we quote this as our best measure of the M87 stellar temperature. However, the shape of the velocity profiles suggests that some parameterization of the LOSVDs may provide useful insight. We have done this and show our results in Figure \ref{fig:fits} for a single Gaussian, Gauss-Hermite, and two-Gaussian fit. We discuss each in turn here.

To explore a single Gaussian fit, we have taken two approaches. The first is to fit a single Gaussian to the velocity profiles and measure the second moment (see the upper two panels in Figure \ref{fig:fits}). The measured velocity dispersion for the R1 and R2 fields is $458 \pm 87$~\kms\ and $631 \pm 98$~\kms\ respectively. The second approach is to force a Gaussian LOSVD during the initial kinematic extraction. This approach returned very similar results to fitting a Gaussian to the LOSVD \emph{after} a non-parametric extraction. As neither of these methods accurately captures the velocity profile asymmetry, we will not discuss the single Gaussian fit further.

The second approach is to fit the velocity profiles with a single Gauss-Hermite polynomial \citep{van93,ger93,bin98}. This sum of orthogonal functions modifies the pure Gaussian profile to better capture profile asymmetries. Gauss-Hermite polynomials are widely used in quantifying the skewness and kurtosis (the $h_3$ and $h_4$ moments) of stellar populations \citep[e.g.,][]{ger93,ems04} and was shown by \citet{van93} to return very minimal correlations in errors between the parameters. The second row of Figure \ref{fig:fits} plots the Gauss-Hermite fits to the R1 and R2 velocity profiles. The values of $v$, $\sigma$, $h_3$ and $h_4$ for the R1 field are $v = 65 \pm 13$~\kms, $\sigma = 456 \pm 37$~\kms, $h_3 = 0.16 \pm 0.03$ and $h_4 = 0.06 \pm 0.02$. For the R2 field these values are $v = 111 \pm 23$~\kms, $\sigma = 604 \pm 45$~\kms, $h_3 = 0.15 \pm 0.05$ and $h_4 = 0.005 \pm 0.003$ respectively. Notice that the values of $\sigma$ for the Gauss-Hermite extraction are similar to the single Gaussian fit.

The third approach we take is to fit the R1 and R2 velocity profiles with two Gaussians. Admittedly the choice to fit the LOSVD with two Gaussians (as opposed to 3, etc.) is arbitrary; by parameterizing the velocity profiles in this way we are assuming, a~priori, the existence of two kinematically distinct and Gaussian distributed populations of stars. Our decision is based, in part, on trying to minimize the number of free parameters in the fits while still capturing the details of the velocity profile shape. The general quality of the fit (lower 2 panels in Figure \ref{fig:fits}) qualitatively supports the presence of no more than two Gaussians. Although the two-Gaussian fit is ad hoc, it both yields very low fit residuals and affords a lower limit to the stellar temperature of M87. These lower limits are robust whether the asymmetry stems from a second kinematic component or simply from noise.

  \begin{center}
\begin{deluxetable*}{ccccccc}
  \tablecaption{Two-Gaussian Velocity Profile Fits
    \label{table:2gauss}}
  \tablehead{
    \colhead{} & \colhead{} & \colhead{Gaussian 1} & \colhead{} & \colhead{} & \colhead{Gaussian 2} & \colhead{} \\
    \colhead{Pointing} & \colhead{Area (\%)} & \colhead{Center (\kms)} & \colhead{Sigma (\kms)} & \colhead{Area (\%)} & \colhead{Center (\kms)} & \colhead{Sigma (\kms)}}
  \startdata
     &  &  &  &  &  &  \\
  R1 & 85.6 & $15 \pm 14$ & $383 \pm 32$ & 14.4 & $991 \pm 35$ & $215 \pm 34$ \\ 
  R2 & 74.3 & $-52 \pm 54$ & $446 \pm 43$ & 25.7 & $979 \pm 128$ & $401 \pm 87$ \\
  \enddata
  \tablecomments{Parameters for the best two-Gaussian fit to the R1 and R2 velocity profile as shown in Figure \ref{fig:fits}. The two ``Area'' columns denote the percent of the normalized total area for each Gaussian component.}
\end{deluxetable*}
  \end{center}

\subsection{Uncertainties in the Kinematics}\label{sec:uncertainties}

In \S~\ref{sec:tests} we discuss various tests to explore whether the asymmetry in the LOSVD is a robust result. These tests explore potential systematics in data calibration and aspects of the kinematic extraction process. In this section we turn to potential sources of uncertainty in the data reduction \emph{before} extracting the kinematics. We describe each in turn here, starting with the calculation of the formal uncertainties on the LOSVDs themselves.

\subsubsection{Monte Carlo Uncertainties}\label{sec:mc}
The uncertainties in our LOSVDs (shown in Figures \ref{fig:spec} and \ref{fig:fits}) come from 100 Monte Carlo realizations of the kinematic extraction process, described in \S \ref{sec:extract}. This procedure begins with the best-fit set of template stars (i.e. the red template fit in Figure \ref{fig:spec}). At each wavelength, and for each Monte Carlo realization, a flux value is drawn at random from a Gaussian distribution of the noise at that wavelength. The mean of the sampled Gaussian distribution is the flux from the best-fit convolved and weighted template spectrum and the standard deviation is set to the mean of the noise. The noise estimate comes from Vaccine, which propagates the noise through the entire reduction, pixel by pixel. A new LOSVD is determined for all 100 realizations and the standard deviation of the 100 values within each of the 29 velocity bins forms the final uncertainty of our best-fit LOSVD.

\subsubsection{Sky Subtraction Uncertainties}\label{sec:skyuncertain}
Sky subtraction is the limiting factor when working at this low surface brightness. This is particularly true when the sky is sampled at a different time then when the data are taken, as is the case here. We have already explored the influence of sky subtraction on both the kinematic extraction \citep[][see also Murphy et al. in preparation]{mur11} and on Lick index line strengths \citep{gre12,gre13}. The methods laid out in those papers were carried out in this work. Namely, a range of scalings were applied to the neighboring sky nods before combining and carried through all subsequent data reduction steps as described in \S~\ref{sec:sky}. This approach provides us with a very heuristic, brute-force method of determining the influence of various sky nod scalings on both the final S/N and extracted kinematics. In both MGA11 and \citet{gre12} we find that for the majority of cases, scaling the sky nods by half of the science exposure time, then coadding, produces the best results. This proves true for the data presented here, with 35 of the 38 science frames returning the best results with equal scaling to the sky nods. This stems in large part from the very good observing conditions the data were taken in. From a visual inspection of Figure \ref{fig:spec}, the lack of any clearly over or under subtracted night sky lines gives qualitative assurance that our sky subtraction is not a systematic problem. Also, the large number of frames we combine has a mitigating affect, as slight over-subtraction and under-subtraction tend to cancel out in the final spectrum.

\subsubsection{Selection of Fibers: Splitting the Kinematics}\label{sec:split}

\begin{figure*}[th]
  \begin{center}
  \vspace{0.22cm}
    \centerline{
      \psfig{file=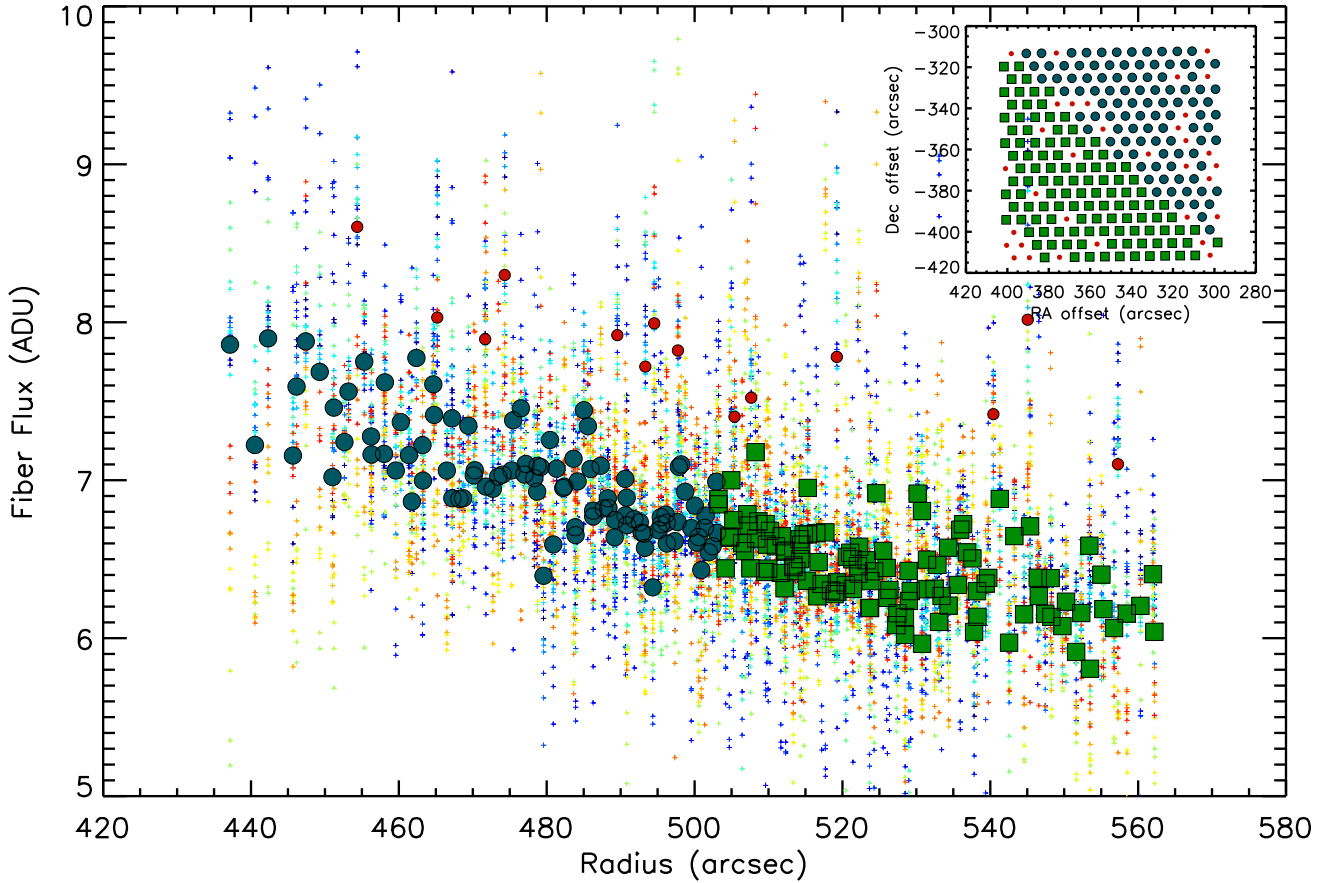,width=18.5cm,bbllx=54bp,bblly=360bp,bburx=558bp,bbury=720bp,clip=}}   
    \figcaption{The median fiber flux as a function of radius. The smaller colored ``$+$'' symbols show the flux values for each individual fiber for each of the 38 science exposures. The heavier filled symbols denote the median of each set of 38 flux values. The large, filled circles (blue) denote the fibers used for the R1 field while the filled squares (green) show fibers combined to form the R2 field. The smaller filled circles (red) plot fibers that show excess flux above the light profile of the galaxy, based on a visual inspection. These fibers were withheld from either bin. The inset to the upper-right shows the on-sky positions of the R1 and R2 fields, and rejected fibers.
    \label{fig:fibflux}}
    \vskip-15pt
  \end{center}
\end{figure*}

Figure \ref{fig:fibflux} plots the median fiber flux as a function of radius from the center of M87 and is used to reject fibers showing evidence for excess flux above the light profile of the galaxy. The small ``$+$'' symbols indicate the flux in each fiber for all 38 exposures. The heavier, filled symbols denote the median of these 38 values. The large (blue) circles indicate all fibers combined in the R1 field. The large (green) squares show those fibers in the R2 field, with smaller (red) circles denoting those fibers withheld from the stacks, based on a visual inspection. The small inset to the upper-right shows the on-sky locations of the R1, R2, and rejected fibers. We explored combining all the fibers, including those showing slightly elevated flux. As the combination of the fiber data is made using the biweight \citep{bee90}, which tends to reject outliers, the fibers showing elevated flux did not alter the final spectrum in a significant way.

Before settling on the choice of fibers in the R1 and R2 fields, we explored a wide range of other fiber combinations. As discussed in \S~\ref{sec:tests} we explored combining all the fibers as well as splitting the data between the May 2010 and March 2011 observing runs. These led to quantitatively similar results. The decision to split the field into the R1 and R2 fields was made to test whether the stark rise in velocity dispersion we measure could be seen to increase across our field. The split was made based on a radial cut of $437\arcsec\ \le \text{R1} \le 506\arcsec\ < \text{R2} \le 562\arcsec$. The location of this division was made to keep the S/N in the R1 and R2 bins comparable. The resulting S/N is 28.9 for the R1 field and 25.2 for the R2 field. As each spectrum is a combination of $\sim 20,000$ pixels at each wavelength, the direct calculation of the noise, based on the pixel-to-pixel noise coming from the Vaccine data reductions is prone to very slight systematics. Therefore, these S/N values are taken from the RMS values of the mismatch between our template stars and the galaxy spectrum fit as described in \S \ref{sec:extract}.

\subsubsection{The Systematic Offset in \mgb}\label{sec:mgb}

In MGA11 (see Figures 5 and 6) we found that our stellar velocity dispersion measured for the \mgb\ region (4930--5545~\AA) were systematically lower than the other 4 spectral regions used to extract kinematics. The systematic offset was $\sim 20$~\kms\ and is in good agreement with the SAURON data which employs a similar wavelength range (4810--5400~\AA). For these data, due to S/N limitations, we have not split our spectral range into discrete wavelength bins as was done in MGA11. Therefore, the \mgb\ region plays a non-negligible role in constraining our extracted LOSVD and subsequent measure of the stellar velocity dispersion. However, we note that if such a systematic is in these data, it drives our reported stellar velocity dispersion measurements to lower values, not higher. Moreover, such an offset still falls within our uncertainty and is mitigated by the other spectral features that fall within our 4100--5400~\AA\ range.

\section{Results}\label{sec:results}

\begin{figure*}[th]
  \begin{center}
  \vspace{0.22cm}
    \centerline{
      \psfig{file=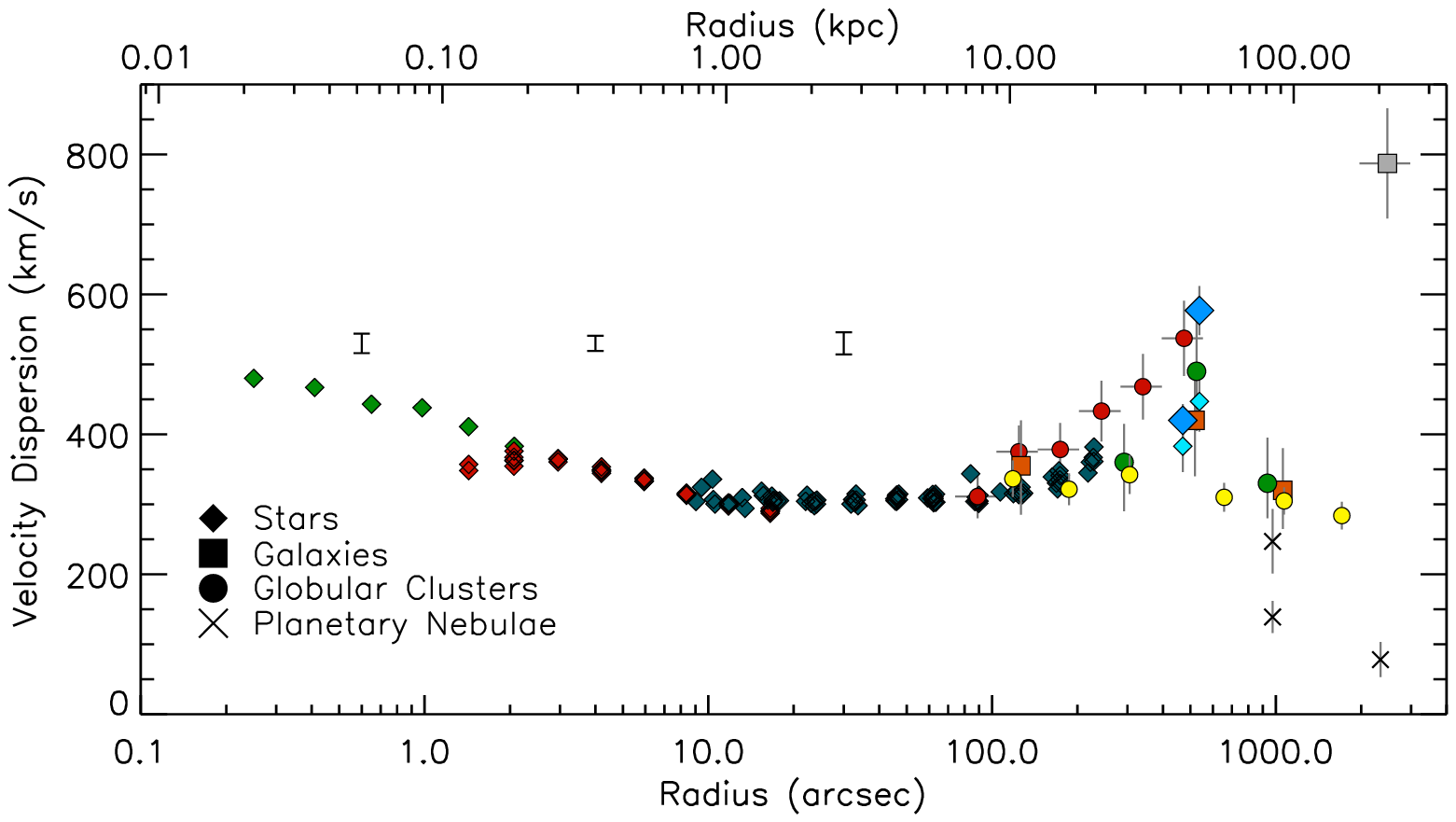,width=18.5cm,bbllx=54bp,bblly=440bp,bburx=558bp,bbury=700bp,clip=}}   
    \figcaption{Velocity dispersion vs. radius for integrated stellar light, globular cluster, UCDs and planetary nebulae data for M87. Each kinematic component is denoted as a different symbol as per the figure legend. The green diamonds plot the stellar velocity dispersion measurements from the NIFS data of \citet{geb11} and span 0.2\arcsec\ to 2\arcsec. The red diamonds plot the publicly available SAURON data \citep{ems04} over the spatial range used in MGA11 (1.2\arcsec\ to 17\arcsec). The Mitchell Spectrograph stellar velocity dispersion values from MGA11 are shown as blue diamonds and extend from 8\arcsec\ to $\sim 240$\arcsec. The individual uncertainties on these data points have been suppressed for visual clarity. However, we plot typical uncertainties directly above each set of data points. The red circles plot the GC kinematic data from \citet{han01}. These data extend from 88\arcsec\ to 475\arcsec\ and are slightly different than the \citet[][see their Table 1]{cot01} velocity dispersion values from the same data; in MGA11 we rebinned the \citet{han01} individual GC velocities to better match our dynamical modeling spatial bins. The yellow circles plot the GC data from S11 which range between $\sim 110$\arcsec\ to $\sim 1580$\arcsec\ and show a clearly different trend in velocity dispersion from the \citet{han01} data, beginning at $\sim 200$\arcsec. The bright GCs from S11 are shown as green circles and their UCDs are plotted as orange squares. The black crosses plot the PNe data from \citet{doh09}, as described in the text. The Virgo Cluster galaxy velocity dispersion, calculated by us from the individual galaxy data of \citet{rin08}, is plotted as a gray square in the far upper-right. Our new stellar non-parametric measurements of the full velocity profiles (right-hand column in Figure \ref{fig:spec}) are denoted as large, blue diamonds at 480\arcsec\ and 526\arcsec. The uncertainties for the new stellar data points come from Monte Carlo simulations as described in \S~\ref{sec:uncertainties}. We also plot the velocity dispersions taken from the primary Gaussian of the two-Gaussian extraction (smaller, light-blue triangles) and consider these lower limits. The trend of an increasing velocity dispersion for the Mitchell Spectrograph stellar kinematics is clear, beginning at $\sim 100$\arcsec\ and continuing to the limits of our spatial coverage.
      \label{fig:disp}}
    \vskip-10pt
  \end{center}
\end{figure*}

In Figure \ref{fig:disp} we plot the velocity dispersion from several sources along with our new stellar velocity dispersion measurements. In the central region (0.2\arcsec\ to 2\arcsec) the green diamonds plot data from \citet{geb11} taken with the Near-infrared Integral Field Spectrograph (NIFS) on the Gemini telescope. The red diamonds plot the publicly available data from SAURON \citep{ems04} and cover 1.2\arcsec\ to 17\arcsec, the same radial range used in MGA11. The blue diamonds plot the Mitchell Spectrograph data from MGA11 and cover $\sim 10$\arcsec\ to $\sim 240$\arcsec. The uncertainties for these three sets of data points have been suppressed for visual clarity, with the average uncertainty shown directly above each data set.

The rise in stellar velocity dispersion begins just beyond 100\arcsec\ and follows the trend seen in the GC data from \citet{han01}, plotted as red circles with uncertainties. These GC velocity dispersion values are different than those in \citet[][see their Table 1]{cot01} who used the Hanes et al. GC velocities for their own dynamical analysis; in MGA11 we elected to rebin the individual GC velocity values from Hanes et al. to better align with our modeling spatial bins. Although the values plotted here are different than those in \cote\ they are similar as the steeply rising trend with radius is the same.

The GC velocity dispersion data from S11 are plotted as yellow circles and agree with the stellar kinematics of MGA11 between $\sim 100$\arcsec\ and $\sim 250$\arcsec. However, at R $\ge 250$\arcsec\ the GC kinematics of S11 begin to strongly diverge from both the GCs of \citet{han01} and our new stellar measurements. We also plot the bright GCs (green circles) and UCDs (orange squares) from S11. The black crossess at large radii plot the \citet{arn04} and \citet{doh09} PNe velocity dispersion values. The two different values at R~=~822\arcsec\ are PNe from \citet{arn04}, with the lower value coming from a reevaluation of the data as presented in \citet{doh09}. In this work, 3 of the 12 PNe are classified as potential intracluster PNe and withheld from the dispersion measurement. Doherty et al. conducted a $\chi^2$ analysis of the 2 data sets but could not distinguish between the two. In either case, these velocity dispersion values are well below all the data plotted in Figure \ref{fig:disp}. Lastly, plotted in the upper-right as a gray square is the velocity dispersion of a set of galaxies in Virgo, taken from the recession velocities of \citet{rin08}. We have calculated both the radial distance to M87 and velocity dispersion from these data, with the uncertainty in radius showing the full radial range of Virgo galaxies used in calculating the velocity dispersion. Rising to a velocity dispersion of nearly 800~\kms, the Virgo Cluster DM halo is certainly the dominant mass component by 200~kpc. The sharp contrast this value makes with the data point from \citet[][]{doh09} at a similar radial extent is intriguing and could indicate the presence of multiple dark matter sub-halos.

The data points we have added are plotted as the large, light-blue diamonds and are the non-parametric measurements of the second moment for the R1 and R2 fields. We also include the dispersion values taken from the primary Gaussian of the two-Gaussian fit (smaller, light-blue triangles), as explained in \S~\ref{sec:fitting}, and are considered lower limits. Of particular note is the good agreement at the R~$\approx 8$\arcmin\ position between the bright GCs, UCDs and our new stellar measurements. We discuss both the asymmetry of the velocity profile, the disagreement with both the GC and PNe values, and the rising velocity dispersion profile of M87 in \S~\ref{sec:discussion}. 

\section {Discussion}\label{sec:discussion}

The asymmetry of the stellar velocity profiles indicates complex stellar dynamics at the position of our pointing. In an analysis of the stellar dynamics we are confronted with two coupled issues. First is the issue of an accurate measurement of the second moment of the stellar component of M87. This measure is complicated by the strong asymmetry in the LOSVD. We have given the non-parametric measurement of the second moment as the most direct interpretation of the velocity profile. The parametric, Gauss-Hermite measure of the velocity profile also returns very high values for the second moment. This brings us to the second issue. That is, the intriguing possibility that the strong asymmetry in the velocity profile stems from a cooler, second component of stars superimposed along the line-of-sight and offset to a positive velocity of $\sim 1000$~\kms\ from the rest velocity of M87. The evidence for a second component is explored in \S~\ref{sec:formation} and \S~\ref{sec:conflict}. For now we simply note that if there is a second component, fitting our velocity profiles with two Gaussians yields a more representative measure of M87's stellar halo temperature. Yet even this most conservative measure of the stellar velocity dispersion ($383 \pm 32$~\kms\ and $446 \pm 43$~\kms) is still well above both the gradually declining GC velocity dispersion values of S11, and the PNe measurements of \citet{doh09}.

What can we make of these conflicting results between the stars, GCs and PNe kinematics? The discrepancies between our stellar velocity dispersion values and both the GC kinematics from S11 and PNe measurements from Doherty et al. are intriguing, but not entirely surprising; there is extensive evidence indicating that the center of the Virgo Cluster is still in active formation \citep{tul84,bin87,bin93,wei97,mih05,doh09,rud10,kra11,rom12}. In considering our results in this light, we believe our pointing falls on a dynamically hot and complex region of M87 and thus reduces the tension between these potentially disparate data sets. We outline some of the relevant work done on the dynamics of the center of M87 in \S~\ref{sec:formation} and explore the tension between these and previous results in \S~\ref{sec:conflict}. Then in \S \ref{sec:rising} we explore the evidence for a rising velocity dispersion and massive dark matter halo.  We then compare the connection between M87 and several BCGs that exhibit a similar rising velocity dispersion profile.

\subsection{A System in Formation}\label{sec:formation}

In the deep photometry of \citet{mih05} there is evidence of elongated and disturbed isophotes towards the SE major axis. Moreover, superimposed on the galaxy light of M87 is extensive intracluster light (ICL) \citep{mih05,jan10,rud10}, formed from stripped stars bound to the Virgo Cluster rather than a specific galaxy. The ICL has been studied extensively over the past decade \citep[for a nice overview, see][]{arn10} and is predicted to be ubiquitous in galaxy clusters \citep[e.g.,][]{wil04}. In \citet{wei97} they find a stellar stream extending to nearly 100~kpc along the SE major axis, directly across our field (see Figure 1 in that work) and speculate this stellar debris comes from the recent accretion of a spheroidal galaxy. With a short dynamical time for the extended material (t~$\le 5 \times 10^8$ years) the authors argue we have either caught M87 during a special time during its formation, or that these events are common in the buildup of the outer halos of massive ellipticals. As recent theoretical work on the mass assembly of ellipticals points to minor mergers as a primary mode of mass assembly \citep{naa09,ose10,joh12}, and that the bulk of the ICL is built by the stripping of stars during the formation of the BCG \citep[e.g.,][]{mur07} it appears likely these events are common.

Further evidence of the continued assembly of M87 comes from the GC kinematical analysis of \citet{rom12} where the phase-space substructure of the M87 GC population reveals the existence of at least 2 components. In their work (see their Figure 1) a shell-like structure is discovered. Their formation simulations seem to indicate the accretion of a $\sim 0.5$~L$^*$ elliptical progenitor with an observable lifetime of $\sim 1$~Gyr. As the authors point out, there is difficulty in reconciling the large number of GCs in the shell component with the relatively cold velocity dispersion they measure. One plausible scenario they suggest is that the accretion of an $\sim$~L$^*$ galaxy (i.e. a $\sim 1:10$ merger) combined with the stripping of GCs from several satellite compact dwarf galaxies could lead to both a large number and simultaneously cool GC component.

Extensive work on the dynamics of the Virgo Cluster as a whole,\footnote{See the introduction in \citet{fou01} for a good overview of the studies of Virgo galaxies.} particularly the pioneering work of Binggeli and collaborators \citep{bin85,bin87,bin93}, has shown clear evidence that the Virgo Cluster is young and unrelaxed. The Virgo cluster is not unique in this regard and signs of continued assembly can be seen in other galaxy clusters such as the Coma Cluster \citep[e.g.,][]{ger07}. Intriguingly, the distribution of Virgo galaxy velocities shows a pronounced asymmetry to positive velocities, with a peak offset by $\sim 1000$~\kms\ from the Virgo recession velocity \citep{bin93,fou01,mei07}.  Although a much more comprehensive dynamical analysis is needed to say anything definitive, the idea that one of these galaxies was stripped during a close passage to M87, depositing a relatively cool stellar component at a velocity offset of $\sim 1000$~\kms, is one possible source of the velocity profile asymmetry we see.

\subsection{Conflicting Results?}\label{sec:conflict}

Neither the GC measurements of S11, nor the PNe measurements of \citet{arn94} and \citet{doh09}, are in line with our stellar velocity dispersion measurements. Moreover, the GC measurements of \citet{han01}, as rebinned for the work of MGA11 and shown in Figure \ref{fig:disp}, are significantly higher than 3 of our 4 stellar measurements. We now turn to each set of conflicting results in search of some relief to this tension in the observations.

As the discrepancy between the PNe measurements and the stars is the most dramatic, we will begin there. Due to the evidence of continued assembly of the Virgo Cluster, we believe the comparison to the PNe data from \citet{doh09} is simply ill-advised; their data comes from the opposite side of the galaxy as our stellar kinematics (see their Figure 7) and is likely dynamically unrelated to the stars in our field. Also of note is the difference in radial position between the PNe and stellar measurements. The Doherty et al. pointings are centered at 13.7\arcmin\ and 32.9\arcmin\ compared to our data points at 8.0\arcmin\ and 8.8\arcmin. When seen in this light, agreement between the PNe and stellar measurements would be more surprising than not.

Next we explore the conflicting results between the GC measurements of S11 and \citet{han01}. As stated earlier, in order to match the dynamical modeling bins of MGA11 we rebinned the individual Hanes et al. GC velocities. These values are plotted in Figure \ref{fig:disp} and are slightly different than what is found in the dynamical analysis of \citet{cot01}. As discussed in S11, the Hanes et al. values were found to contain a handful of ``catastrophic outliers''. Once these outliers are removed, and their new GC data included, their measured dispersion drops substantially. This relieves the tension between the GC measurements, but still does not explain why our stellar velocity dispersion values are significantly higher than the S11 GC values. Unlike the PNe values, the spatial agreement between the S11 GCs and our field is good; we have 4 of their GCs within 200\arcsec\ from the center of our field. These coincident GCs are very cold and bare no resemblance to the dynamics we see in the stars.\footnote{The GCs recession velocities (relative to the rest velocity of M87) within 200\arcsec\ of our field from S11 are as follows. S87: $83\pm103$~\kms, S93: $-12\pm48$~\kms, S170: $-22\pm106$~\kms, S270: $86\pm43$~\kms.} However, this apparent conflict between the stars and GC kinematics of S11 gets some relief by taking a step back and considering the GC population of M87 as a whole. 

There are a couple of indications that the 8\arcmin~$\le$~R~$\le$~10\arcmin\ region of M87 is dynamically hotter and more complicated than other regions of the galaxy. Figure 23 in S11 shows the GC LOSVDs for the 4 subpopulations they define in their sample. In their inner radial bin (R~$\le$~10\arcmin) both the blue GCs, bright GCs (i$_0 < 20$) and UCDs show distinctive wings to positive velocity, similar to the LOSVDs of the stars in this work. The wings to positive velocity are also seen in the GC velocity profiles of \citet[][Figure 2]{rom01}. Although the agreement between the degree of offset in the velocity profile asymmetry seen in the S11 GCs and our stars is not perfect, it suggests a possible link between the mechanisms responsible for assembling the blue GCs, bright GCs, UCDs, and stars at this radius. This connection between the stars and blue GCs is also seen in the spatial distribution and chemical analysis of \citet{for12}.

In considering the possibility of such a link more closely, Figure 21 in S11 shows the position angle (PA), rotation velocity, velocity dispersion and velocity kurtosis of their 4 subpopulations of GCs. There are two points we make here. First, the blue GC population (left-most column) shows a distinct change in PA just beyond R~$> 10$\arcmin. This change in PA corresponds to a spike in rotational velocity while the velocity dispersion drops at $\sim 10$\arcmin\ from $\sim 370$~\kms\ (in rough agreement with our two-Gaussian measurement) to $\sim 300$~\kms. Second, of keen interest to us are the UCDs and bright GCs in their sample (right-most column in their Figure 21) which show a strong rise in velocity dispersion to well above 400~\kms\ at $\sim 8$\arcmin. At the same radius the bright GCs reach nearly 500~\kms, and both the UCDs and bright GCs show a noticeable spike in rotation velocity at this position. This spike in velocity dispersion is also shown in the left-hand plot of their Figure 20 where \emph{both} their faint and bright GCs become quite hot, right at R~$\approx 8$\arcmin. These signatures of a change in the dynamical nature of M87 were noted by S11 where they suggest that the 4 dwarf ellipticals found between 7\arcmin\ and 9\arcmin\ are ``stirring the pot''.

This leads us to explore the UCD population as not only the cause of the high stellar velocity dispersion, but also the source of the velocity profile asymmetry. M87 has a radial velocity of 1307~\kms\ \citep[][see also Mei et al. 2007 and Makarov et al. 2011]{huc12}. The UCDs nearest our field are NGC~4486a and IC~3443 (see Figure \ref{fig:mihos}). The radial velocity of NGC~4486a is 757~\kms\ \citep{pru11} and therefore can not be the source of the positive velocity asymmetry. A radial velocity of 2272~\kms\ was reported for IC~3443 in \citet{dev91} which was in good agreement with the earlier work of \citet[][]{eas78} who measured a value of 2254~\kms. At these values, the halo of IC~3443 would appear to be a candidate for the source of the $\sim 1000$~\kms\ wing to positive velocity seen in our velocity profiles. However, more recent radial velocity measurements for IC~3443 return lower values, typically under 2000~\kms\ \citep{bot88,van00,gav04}, with 1785~\kms\ being the currently accepted value \citep{ade06,rin08,aih11}. At this radial velocity, stars from the halo of IC~3443 can not explain the velocity profile wing.
 
In order to determine if there is any photometric evidence for a second population of stars within our field we have inspected the deep photometry of \citet{mih05} to look for any variation beyond the smooth decline in the surface brightness of M87. Despite the aforementioned evidence for disturbed isophotes along the SE edge of the major axis \citep{wei97,mih05}, we see no distinct changes across our field. We also find no significant deviations from a smooth decline in the surface brightness with radius from an inspection of the flux in our fibers shown in Figure \ref{fig:fibflux}. However, despite not finding evidence for a kinematic disturbance in the photometry, \citet{rom12} point out in their discussion that M87 is a favored target for its proximity and ``general lack of obvious dynamical disturbance [sic]''. Yet dynamical studies have now revealed rich phase-space complexity in not only M87 but several other massive ellipticals that is not apparent in the photometry alone. As our new data points exhibit kinematic complexity not seen in the velocity profiles from MGA11 it leads us to wonder whether the outshirts of the stellar halo of M87 hold a wealth of clues to its formation. Clearly, further observations of the stellar halo of M87 are warranted.

\subsection{A Rising Stellar Velocity Dispersion}\label{sec:rising}

We have quoted 3 different estimates of the stellar velocity dispersion of M87 for our R1 and R2 fields: a non-parametric measure of $420 \pm 23$~\kms\ and $577 \pm 35$~\kms, a Gauss-Hermite parameterization of $456 \pm 37$~\kms\ and $604 \pm 45$~\kms, and a two-Gaussian parameterization of $383 \pm 32$~\kms\ and $446 \pm 43$~\kms. We advocate the non-parametric measure as the most straightforward interpretation and also the most relevant for dynamical modeling. However, due to the intriguing, albeit tentative, evidence for the existence of a second stellar component, we remain somewhat agnostic about which fit better represents the true velocity dispersion of the stars in the M87 halo. That said, \emph{even the lowest estimate of the dynamical temperature of the stars at R~$\approx$~500\arcsec\ shows a clear increase to above 400~km~s$^{-1}$.} What can we make of this continued increase in stellar velocity dispersion out to nearly 45~kpc in M87, and how does it inform our picture of galaxy structure and formation?

\subsubsection{The Dark Matter Halo of M87}\label{sec:dmhalo}

In terms of galaxy structure, if the rising velocity dispersion profile reflects the gravitational potential of the galaxy, then the presence of a massive dark matter (DM) halo can explain such a rise. Indeed, a very massive DM halo has been detected in M87 by several groups employing a variety of methods \citep{fab83,huc87,mou87,mer93,rom01,mat02,das10,mur11}. Yet clearly the velocity dispersions shown in Figure \ref{fig:disp} can not all be reflections of the gravitational potential of M87. One relevant point here is that in the case of both GC and PNe measurements one is working with individual data points.  While the GC population of this and other massive early-type galaxies can reach several thousand \citep[e.g.,][]{mcl99}, measurements of their kinematics are typically done with a few hundred GCs. In the work of \citet{doh09}, their PNe measurement of velocity dispersion comes from 12 PNe for the 247~\kms\ measurement and 9 for the lower 139~\kms\ value. S11 are in a far better position, having velocity measurements of over 700 GCs. With the kinematic complexity of M87, sampling sub-populations that are not reflective of the gravitational potential of the galaxy is a possibility. With integrated starlight we avoid these statistical challenges. Yet if the 8\arcmin~$\le$~R~$\le$~10\arcmin\ region is dynamically hotter, and with our much smaller field-of-view than covered by either the GC or PNe populations, we find ourselves in a similar position of not fully sampling the phase space. Certainly more data and a more complete dynamical analysis is in order, but for now we proceed to interpret these results with the understanding that more stellar data is required.

Further support for the presence of a massive DM halo comes from the X-ray gas mass estimates of \citet[][]{mat02}. Their mass estimates align well the with \citet{coh97} GC velocity dispersion profiles (see their Figure 21) which were compiled into the Hanes et al. values. As larger radial data points provide greater leverage on the total enclosed mass of a galaxy, the GC values of Hanes et al. play a significant role in constraining the DM halo measured in MGA11. In that work we plot a comparison of our best enclosed-mass profile to a variety of literature values (Figure 11 and Table 4 of MGA11).  The MGA11 enclosed mass estimate for either an NFW or cored-logarithmic DM halo are generally in good agreement with the mass estimates from the literature, particularly at large radii \citep[e.g.,][]{fab83}.

However, S11 conduct a similar comparison (see their Figure 16) and find a substantially less massive DM halo for M87. This lower mass is certainly driven by their lower velocity dispersion.  As those authors point out, comparisons between data sets from various groups is challenging, and perhaps attempting to align the various mass tracer populations in M87 is ill-advised; the kinematics of M87, and the center of the Virgo Cluster, are complicated and there is no a~priori reason that the GC and stellar kinematics must go in lock step as different formation pathways will leave different kinematic signatures. For example, the accretion of smaller satellite galaxies with cooler kinematic components can lead to a lower velocity dispersion in the GCs than those formed in~situ, as pointed out in \citet[][]{rom12}. These complications necessitate both a more complete set of observations of the kinematic components and comprehensive dynamical modeling in order to get the entire picture of M87's formation history. Yet with the good agreement between the gas kinematics of \citet[][]{mat02} and the stellar kinematics of MGA11 and those presented here, it appears M87's DM halo is the dominant mass component by 30~kpc.

\subsubsection{M87 as the BCG of Virgo?}\label{bcgs}

How does M87 compare to other central galaxies? Although M87 is not technically the BCG of Virgo\footnote{M49 is slightly more luminious and thus the ``brightest'' galaxy in the Virgo Cluster \citep{kor09}.} it does occupy a central location in terms of cluster mass \citep[e.g.,][]{boh94}. We therefore compare M87 to other BCGs and find that rising stellar velocity dispersions are not uncommon.

Early work by \citet{dre79} on the BCG in the Abell~2029 galaxy cluster (IC~1101) found a stellar velocity dispersion that increases to over 500~\kms\ at 100~kpc. \citet{kel02} find a similar result for NGC~6166, the BCG in Abell~2199 \citep[see also][]{car99}, where the stellar velocity dispersion gradually rises to $\sim 660$~\kms\ by 60~kpc. Also, from stellar velocity dispersion measurements and strong lensing constraints, \citet{new11} find the BCG in Abell~383 to exhibit a steeply rising dispersion profile that climbs from $\sim 270$~\kms\ at the center of the galaxy to $\sim 500$~\kms\ by $\sim 22$~kpc. And the extensive work of \citet[][]{lou08} on 41 BCGs with long-slit data find a significant fraction of their sample exhibit flat to rising velocity dispersion profiles.\footnote{In their paper Loubser et al. claim to find 5 galaxies with rising stellar velocity dispersions. However, if we allow for even a gradual rise in velocity dispersion and inspect their figures, we find this number increases to $\sim 14$.} Obviously, rising velocity dispersions in other galaxies do not give support for a rising velocity dispersion in M87. We simply want to highlight that as observations improve, and dynamical measurements at ever-larger radii become possible, we have been finding rising velocity dispersion profiles in BCGs in greater abundance.

This leads us to a final comparison between M87 and another BCG. In our non-parametric measure of the second moment, we see a rise of $\sim 150$~\kms\ over 4.3~kpc. This sharp rise in velocity dispersion over a relatively short radial distance is quite striking, but not unprecedented. In \citet{ven10} the stellar kinematics of the BCG NGC~3311 are found to rise very rapidly from $\sim 150$~\kms\ at the center to $\sim 450$~\kms\ at R~$\approx~13$~kpc. Over a very similar physical distance (from R~$\approx 8$~kpc to R~$\approx 13$~kpc, a difference of $\sim 5$~kpc) the velocity dispersion rises from around 280~\kms\ to above 450~\kms, a steeper rise than we see for M87. \citet{ric11} remeasure the stellar velocity dispersion of NGC~3311 and use GC velocities to constrain the DM halo, and although they report considerably lower values for the overall stellar velocity dispersion than \citet{ven10}, they see the same steep rise over a similar radial distance. Also noteworthy in NGC~3311 is the difference between the stars and GCs; the GC velocity dispersion at R~$\approx~15$~kpc is higher than their stellar velocity dispersion measurements by $\sim 180$~\kms\ \citep[see Figure 4 in][]{ric11}. Both M87 and NGC~3311 remind us of the need for caution when using dynamical tracers to constrain the mass of a galaxy.

\section{Summary}\label{sec:summary}

We have carried out a dynamical analysis on the stars along the SE major axis at R~$\approx 40$ to 45~kpc from the center of M87. Even our most conservative interpretation finds the stellar velocity dispersion exceeding 400~\kms. The stellar kinematics are complex, with a strong asymmetry to positive velocity. Although speculative, the asymmetry may stem from a cooler, second component of stars superimposed on our field.

All of this raises the question of what a rising velocity dispersion tells us about a galaxy. Is it a true reflection of the gravitational potential of the galaxy, the center of the galaxy cluster, or simply a snapshot of a dynamical system that has not reached equilibrium? The question of what our measurements are telling us about a given galaxy becomes more acute when the various dynamical tracers do not agree, as we have found in M87. Although groups have found good agreement between various tracers of mass in a wide range of galaxies \citep[e.g.,][]{coc09,mcn10}, we now know of many galaxies that show strong disagreement, and there is no a~priori reason that the GC, PNe and stellar kinematics must align at all radial positions. We know from simulations that the formation of BCGs appears very active \citep[e.g.,][]{del07a,rus09} and so this type of substructure and disagreement between different dynamical tracers should not be surprising and perhaps even expected.

With a single field at this distance from the center of M87 we can not say much about the overall formation history of M87 beyond confirming that the center of the Virgo Cluster is still undergoing active assembly and the stellar halo exhibits a rising velocity dispersion profile. Further work on the observational front in necessary, yet the ability to constrain the dynamical state of the stars at these unprecedented radial distances points the way towards future exploration. In particular, a larger number of measurements of the kinematics of integrated starlight, taken at a broad range of locations, would be highly illuminating.

\begin{acknowledgments}

JDM is supported by an NSF Astronomy and Astrophysics Postdoctoral Fellowship (AST-1203057) and acknowledges a UT Continuing University Fellowship that helped support him throughout this work. KG acknowledges support from NSF-0908639. The authors would like to thank Chris Mihos for the use of his data and very helpful comments and insights. We also thank The Cynthia and George Mitchell Foundation for their generous support which made the fabrication of the Mitchell Spectrograph possible. Special thanks to Gary J. Hill and Phillip MacQueen on their continued work to make the Mitchell Spectrograph such a successful instrument. JDM thanks Jenny E. Greene for very constructive comments on the manuscript, and the Physics Department at Southwestern University for the funds to allow Mason Cradit to participate in observations of these data. JDM also thanks Joshua J. Adams for discussions regarding subtle points of the sky subtraction routines implemented in Vaccine. The authors also thank Dave Doss, Kevin Meyer, Brian Roman, John Kuehne, Coyne Gibson and all of the staff at McDonald Observatory who helped immensely with the successful collection of these data and the day-to-day handling of the telescope and instrument. This research has made use of the NASA/IPAC Extragalactic Database (NED) which is operated by the Jet Propulsion Laboratory, California Institute of Technology, under contract with the National Aeronautics and Space Administration. 

\end{acknowledgments}

\newpage

\nocite{mei07,mak11,pog13}

\bibliographystyle{apj}
\bibliography{all}

\end{document}